\documentclass[11pt]{article}
\usepackage{moriond,epsfig}

\def\Journal#1#2#3#4{{#1} {\bf #2}, #3 (#4)}

\def\be{\begin{equation}}
\def\ee{\end{equation}}
\def\bea{\begin{eqnarray}}
\def\eea{\end{eqnarray}}

\begin{document}
\vspace*{4cm}
\title{THREE TESTS OF $\Lambda$CDM}

\author{ C. J. A. P. MARTINS }

\address{Centro de Astrof\'{\i}sica, Universidade do Porto, Rua das Estrelas, 4150-762 Porto, Portugal}

\maketitle\abstracts{
The observational evidence for the acceleration of the universe demonstrates that canonical theories of gravitation and particle physics are incomplete, if not incorrect. The next generation of astronomical facilities must both be able to carry out precision consistency tests of the standard cosmological model and search for evidence of new physics beyond it. I describe some of these tests, and discuss prospects for facilities in which the CAUP Dark Side team is involved, specifically ESPRESSO, Euclid and CODEX. }

\section{The dark side of the universe}
In the middle of the XIX century Urbain Le Verrier and others mathematically discovered two new planets by insisting that the observed orbits of Uranus and Mercury agreed with the predictions of Newtonian physics. The first of these---Neptune---was soon observed by Johann Galle and Heinrich d'Arrest. However, the second (dubbed Vulcan) was never found. The discrepancies in Mercury's orbit were a consequence of the fact that Newtonian physics can't adequately describe Mercury's orbit, and accounting for them was the first success of Einstein's General Relativity.

Over the past several decades, cosmologists have mathematically discovered two new components of the universe---dark matter and dark energy---which have so far not been directly detected. Whether the will prove to be Neptunes or Vulcans remains to be seen but even their mathematical discovery highlights the fact that the standard $\Lambda$CDM paradigm, despite its phenomenological success, is at least incomplete.

Something similar applies to particle physics, where to some extent it is our confidence in the standard model that leads us to the expectation that there must be new physics beyond it. Neutrino masses, dark matter and the size of the baryon asymmetry of the univserse all require new physics, and---significantly---all have obvious astrophysical and cosmological implications. Further progress in fundamental particle physics will increasingly depend on progress in cosmology.

One must therefore carry out explicit consistency tests of the standard cosmological model and search for evidence of new physics beyond it. For example fundamental scalar fields are crucial in the standard particle physics model (cf. the Higgs field) and are also invoked in several key cosmological contexts, including inflation, cosmological phase transitions and their relics (cosmic defects), dynamical dark energy powering the current acceleration phase, and varying fundamental couplings. Even more important than each of these paradigms is the fact that they don't occur alone: this will be crucial for future consistency tests.

\section{Varying fundamental couplings}

Nature is characterized by a set of physical laws and fundamental dimensionless couplings, which historically we have assumed to be spacetime-invariant. For the former this is a cornerstone of the scientific method, but for the latter it is only a simplifying assumption without further justification. These couplings determine the properties of atoms, cells, planets and the universe as a whole, so it's remarkable how little we know about them---we have no 'theory of constants'. If they vary, all the physics we know is incomplete. Such a detection would be revolutionary, but even improved null results are important and useful: natural scale for cosmological evolution would be Hubble time, but current bounds are 6 orders of magnitude stronger~\cite{Rosenband}.

Recent astrophysical evidence from quasar absorption systems~\cite{Webb} suggests a parts-per-million spatial variation of the fine-structure constant $\alpha$ at low redshifts; although no known model can explain such a result without considerable fine-tuning, there is also no identified systematic effect that can explain it. One possible cause for concern (with these and other results) is that most of the existing data has been taken with other purposes, whereas this kind of neasurements needs customized analysis pipelines~\cite{Rodger}. An ongoing UVES Large Programme dedicated to these tests should soon provide and independent test \cite{Molaro}. In the longer term a new generation of high-resolution, ultra-stable specrographs like ESPRESSO and CODEX will significantly improve the precision of these measurements.

At much higher redshifts, the CMB is an ideal, clean probe of varying $\alpha$, which will impact the ionization history of the universe (energy levels and binding energies are shifted, and the Thomson cross-section is proportional to $\alpha^2$).
Having said this, bounds are relatively weak due to degeneracies, and the percent barrier has only recently been broken~\cite{Eloisa09}. In any realistic model where $\alpha$ varies other couplings are also expected to vary, and such coupled variations can also be constrained~\cite{Martins10}. One can also constrain the coupling between the putative scalar field and electromagnetism, independently (and on a completely different scale) from what is done in local tests~\cite{Calabrese}.

The recent CMB measurements from WMAP and arcminute angular scales (from ACT and SPT) suggest that the effective number of relativistic degrees of freedom is larger than the standard value of $N_{\rm eff} = 3.04$, and inconsistent with it at more than two standard deviations. We have recently shown~\cite{Eloisa12} that if one assumes this standard value this CMB data significantly improves previous constraints on ${\alpha}$, with ${\alpha}/{\alpha}_0 = 0.984 \pm 0.005$, i.e. hinting also to a more than two standard deviation from the current, local, value. A significant degeneracy is present between ${\alpha}$ and $N_{\rm eff}$, and when variations in the latter are allowed the constraints on ${\alpha}$ are consistent with the standard value. Again it's worth stressing that deviations of either parameter from their standard values would imply the presence of new, currently unknown physics. 

Many compact astrophysical objects can also be used to search for spacetime variations of fundamental couplings, including
Population III stars~\cite{Ekstrom}, neutron stars~\cite{Angeles} and solar-type stars~\cite{Vieira}.

\section{Dynamical dark energy}

Observations suggest that the universe is dominated by component whose gravitational behavior is similar to that of a cosmological constant. Its value is so small that a dynamical scalar field is arguably more likely. Such a field must be slow-rolling (which is mandatory for $p<0$) and be dominating the dynamics around the present day. It follows~\cite{Carroll} that couplings of this field to the rest of the model (which will naturally exist, unless an ad hoc symmetry is postulated to suppress them) lead to potentially observable long-range forces and time dependencies of the constants of nature.

Standard observables such as supernovae are of limited use as dark energy probes~\cite{Maor}. A clear detection of varying $w(z)$ is key, since $w\sim-1$ today. Since the field is slow-rolling when dynamically important, a convincing detection of $w(z)$ will be tough at low redshift, and we must probe the deep matter era regime, where the dynamics of the hypothetical scalar field is fastest. Varying fundamental couplings are ideal for probing scalar field dynamics beyond the domination regime~\cite{Nunes}. We have recently shown~\cite{Amendola} that CODEX can constrain dark energy better than supernovae (its key advantage being huge redshift lever arm), and even ESPRESSO can provide a significant contribution.

Dark energy reconstruction using varying fundamental constants requires an assumption on the field coupling, but there are in-built consistency checks, so that inconsistent assumptions can be identified and corrected~\cite{Pauline}. On the other hand this analysis allows scalar field couplings to be measured and compared to local constraints. Interesting synegies also exist between these ground-based spectroscopic methods and Euclid, which need to be further explored.

\section{The quest for redundancy}

Whichever way one finds direct evidence for new physics, it will only be believed once it is seen through multiple independent probes. This was manifest in the case of the discovery of the recent acceleration of the universe (where the supernova results were only accepted by the wider community once they were confimed through CMB, large-scale structure and other data) and it is clear that history will repeat itself in the case of varying fundamental couplings and/or dynamical dark energy. It is therefore important to develop consistency tests---in other words, astrophysical observables whose behaviour will also be non-standard as a consequence of either or both of the above.

The temperature-redshift relation, $T(z)=T_0(1+z)$, is a robust prediction of standard cosmology; it assumes adiabatic expansion and photon number conservation, but is violated in many scenarios, including string theory inspired ones. At a phenomenological level one can parametrize deviations to this law by adding an extra parameter, say $T(z)=T_0(1+z)^{1-\beta}$. Measurements of the SZ effect at resdshifts $z<1$, combined with spectroscopic measurements at redshifts $z\sim2-3$ yield the direct constraint $\beta=-0.01\pm0.03$~\cite{Noterdaeme}. Forthcoming data from Planck, ESPRESSO and CODEX will lead to much stronger constraints \cite{Atrio,Gemma}.

The distance duality relation, $d_L=(1+z)^2d_A$, is a robust prediction of standard cosmology; it assumes a metric theory of gravity and photon number conservation, but is violated if there's photon dimming, absorption or conversion. At a phenomenological level one can parametrize deviations to this law by adding an extra parameter, say $d_L=(1+z)^{2+\epsilon}d_A$. In this case, current constraints are $\epsilon=-0.04\pm0.08$~\cite{Avgoustidis}.

In fact, in many models where photon number is not conserved the temperature-redshift relation and the distance duality relation are not independent. With the above parametrizations one can show~\cite{Gemma} that $\beta=-2/3\epsilon$, but in fact a direct relation exists for any such model, provided the dependence is in redshift only (models where there are frequency-dependent effects are more complex). This link allows us~\cite{Gemma} to use distance duality measurements to further constrain $\beta$, and we recently found $\beta=0.004\pm0.016$ up to a redshift $z\sim 3$, which is a $40\%$ improvement on the previous constraint. With the next generation of space and ground-based experiments, these constraints can be further improved by more than one order of magnitude. 

\section{Outlook}

Observational evidence for the acceleration of the universe demonstrates that canonical theories of cosmology and particle physics are incomplete, if not incorrect. Several few-sigma hints of new physics exist, but so far these are smoke without a smoking gun; it's time to actively search for the gun.

The forthcoming generation of high-resolution ultra-stable spectrographs will play a key role in this endeavour, by enabling a new generation of precision consistency tests of the standard cosmological paradigm and its extensions. Further exciting possibiblites, not explicitly discussed in this contribution, include direct astrophysical Equivalence Principle tests and E-ELT measurements of the redshift drift (on the latter, see Pauline Vielzeuf's contribution to these proceedings). Finally, there are interesting synergies with space facilities, particularly Euclid, which should be further studied.

\section*{Acknowledgments}
The work of CJM is supported by a Ci\^encia2007 Research Contract, funded by FCT/MCTES (Portugal) and POPH/FSE (EC), with further support from grant PTDC/CTE-AST/098604/2008. Many interesting discussions with the rest of CAUP's Dark Side team and our collaborators elsewhere have shaped my views on this subject, and are gratefully acknowledged.

\end{document}